# Heat dissipation in few-layer MoS$_2$ and MoS$_2$/hBN heterostructure

Alois Arrighi[1,2]*, Elena del Corro[1], Daniel Navarro Urrios[3], Marius V. Costache[1], Juan F. Sierra[1], Kenji Watanabe[4], Takashi Taniguchi[4], J.A. Garrido[1,5], Sergio O. Valenzuela[1,5]*, Clivia M. Sotomayor Torres[1,5] and Marianna Sledzinska[1]*

[1] Catalan Institute of Nanoscience and Nanotechnology (ICN2), CSIC and BIST, Campus UAB, Bellaterra, 08193 Barcelona, Spain
[2] Departamento de Física, Universidad Autónoma de Barcelona, Bellaterra, E-08193 Barcelona, Spain
[3] MIND-IN2UB, Departament d'Enginyeria Electrònica i Biomèdica, Facultat de Física, Universitat de Barcelona, Martí i Franquès 1, 08028 Barcelona, Spain
[4] Advanced Materials Laboratory, National Institute for Materials Science (NIMS), 1-1 Namiki, Tsukuba, 305-0044, Japan
[5] ICREA, Pg. Lluís Companys 23, 08010 Barcelona, Spain

*Corresponding authors: alois.arrighi@gmail.com, SOV@icrea.cat, marianna.sledzinska@icn2.cat

## Abstract

State-of-the-art fabrication and characterisation techniques have been employed to measure the thermal conductivity of suspended, single-crystalline MoS$_2$ and MoS$_2$/hBN heterostructures. Two-laser Raman scattering thermometry was used combined with real time measurements of the absorbed laser power, which allowed us to determine the thermal conductivities without any assumptions. Measurements on MoS$_2$ layers with thicknesses of 5 and 14 nm exhibit thermal conductivity in the range between 12 and 24 Wm$^{-1}$K$^{-1}$. Additionally, after determining the thermal conductivity of a selected MoS$_2$ sample, an hBN flake was transferred onto it and the effective thermal conductivity of the heterostructure was subsequently measured. Remarkably, despite that the thickness of the hBN layer was less than a third of the thickness of the MoS$_2$ layer, the heterostructure showed an almost eight-fold increase in the thermal conductivity, being able to dissipate more than 10 times the laser power without any visible sign of damage. These results are consistent with a high thermal interface conductance $G$ between MoS$_2$ and hBN and an efficient in-plane heat spreading driven by hBN. Indeed, we estimate $G$ ~70 MW·m$^{-2}$K$^{-1}$ which is significantly higher than previously reported values. Our work therefore demonstrates that the insertion of hBN layers in potential MoS$_2$–based devices holds the promise for efficient thermal management.

Keywords: MoS$_2$, hexagonal BN, 2D heterostructure, thermal conductivity, thermal interface conductance

## Introduction

Molybdenum disulfide, MoS$_2$, arguably the most studied 2D material (2DM) after graphene, has attracted much interest due to its semiconducting nature, high carrier mobility and tunable bandgap, which are highly relevant for potential electronic and optolectronic devices[1,2]. However, as high-performance devices have been demonstrated in research laboratories, it remains to be addressed how they would respond in industrial implementations, for which thermal dissipation plays a key role. In this context, MoS$_2$ thermal conductivity measurements have yielded relatively small values, scattered over $k_{MoS_2}$ ~15 to 100 Wm$^{-1}$K$^{-1}$ [3–9]. These results have two important implications. First, they highlight the relevance of proper thermal management as the low $k_{MoS_2}$ can impair the performance of MoS$_2$ in, for instance, field effect transistors (FETs). Because the heat cannot be efficiently dissipated from the device active area solely by MoS$_2$, the thermal interface conductance $G$ between MoS$_2$ and its surroundings acquires special relevance. Second, it is necessary to implement techniques to accurately determine the thermal conductivity of 2DMs and establish the origin of the discrepancies observed in the literature. The spreading in reported results is not only limited to MoS$_2$ as thermal conductivity measurements of 2DMs are challenging from the perspectives of material growth and device fabrication. However, while the spread could partly be attributed to different sample preparation methods and thickness, it has been argued that a more plausible reason is measurement errors[10].

Raman scattering-based methods have been widely used to determine the thermal conductivity of 2DMs[3,4,6,7,11,12]. In one-laser Raman thermometry (1LRT), a laser is used both as a heater and temperature probe (Fig. 1a) and the thermal conductivity is obtained from the temperature rise and the power absorbed [3,4,6,11]. The main drawback of this method is that only the temperature at the center of the sample is known with information neither on the temperature decay along the sample nor at its edges, in contact with the heatsink. In fact, in the specific case of the sample being in thermal equilibrium, there are several sources of error, including the interaction between the 2DM and the substrate, the role of the substrate as heat-sink, the temperature-dependent Raman scattering frequency shift calibration, the environment (*e.g.* air, vacuum) and the determination of the absorbed laser power, which can vary with temperature[10]. Such limitations are even more critical in multilayer heterostructures, where the power, partially absorbed in each layer, has to be assumed, leading to additional uncertainty in the results.

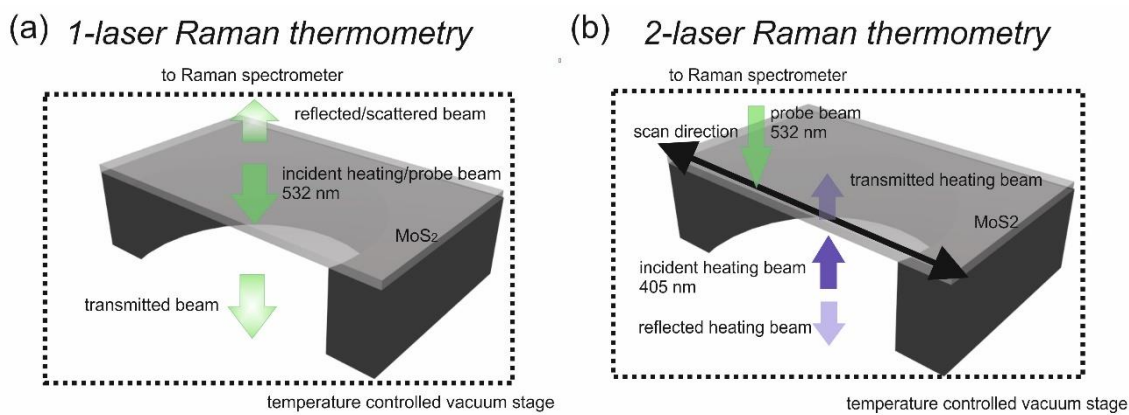

**Figure 1.** Schematics of experimental set-up in (a) one-laser Raman thermometry (1LRT) (b) two-laser Raman thermometry (2LRT) configuration.

Here, we overcome these limitations by implementing two-laser Raman thermometry (2LRT) to determine the thermal conductivity of suspended, single-crystalline MoS$_2$ layers and a MoS$_2$/hBN heterostructure. As the 2D layers are free-standing, errors deriving from thermal interface effects with the substrate are eliminated, thus simplifying data analysis. In contrast to 1LRT, in 2LRT one laser creates a hotspot at the center of the sample, whereas the other laser directly probes the generated temperature profile (Fig. 1b). In this way, the thermal conductivity of the material can be extracted without any assumptions. The 2LRT technique has been successfully used to study silicon nanomembranes[13], phononic crystals[14–16], thermal diodes[17] and polycrystalline $MoS_2$[18]. However, no results have yet been reported using this technique for crystalline 2DMs, perhaps due to the inherent difficulty of fabricating large enough suspended samples. For the heterostructure, we chose hBN because it has been shown to dramatically increase the $MoS_2$ carrier mobility. Furthermore, hBN is an excellent 2D dielectric material with a high reported thermal conductivity ($k_{hBN} \sim$ 200-800 $Wm^{-1}K^{-1}$ [19–21]). The use of hBN as a thermal interface material in 2D devices can thus improve thermal management and overall lead to increased device performance, with higher carrier mobility and thermal dissipation in FETs [22,23].

In order to disentangle $G$ from the thermal conductivities of each layer in the $MoS_2$/hBN heterostructure, $k_{MoS_2}$ and $k_{hBN}$, we have implemented a singular experimental protocol. We first determine $k_{MoS_2}$ on a suspended $MoS_2$ flake using 2LRT. Then a thin hBN layer is selected and transfer onto the already characterized $MoS_2$. By measuring the thermal conductivity of the bilayer with 2LRT, we can extract $k_{hBN}$ and the thermal interface conductance $G$. As hBN is transparent at the wavelength of the used laser, the laser power is mainly absorbed by $MoS_2$. Therefore, the $MoS_2$ temperature rise for a given laser power can be demonstrated to be largely determined by $G$, which is estimated to be $\sim$70 $MW \cdot m^{-2}K^{-1}$, depending on the assumed $k_{hBN}$. This result is significantly larger than previously reported values and opens the way for thermal management using van der Waals heterostructures.

**Experimental details**

In order to support the free-standing 2DM layers, silicon membranes with a thickness of 3 μm were fabricated from a silicon-on-insulator wafer using standard microfabrication techniques. Holes with diameters of 20 μm were then patterned on the membranes using a focused ion beam. The hole-size was selected to be large enough to obtain sufficiently detailed temperature maps of the suspended layers. The 2DMs were mechanically exfoliated using viscoelastic polydimethylsiloxane (PDMS) films. Selected flakes were located with an optical microscope and then transferred onto the holey membranes using all dry transfer methods to avoid contamination[24].

As the Raman-active mode frequency is used as a temperature probe, it is crucial to calibrate its dependence as a function of temperature ($\chi$). During calibration, the entire sample is heated uniformly in vacuum and spectra are recorded at low enough laser power to avoid inducing extra heating. The coefficient $\chi$, known as the first-order temperature coefficient of a given mode, was measured directly in the suspended samples for both $A_{1g}$ and $E_{2g}$ modes (see Supplementary Information).

Two single crystalline $MoS_2$ samples with thickness of 5 and 14 nm have been characterized, which roughly correspond to 7 and 20 monolayers, respectively (see Supporting Information). These samples were used to establish the implementation of 2LRT in single crystalline 2DM. After obtaining $k_{MoS_2}$ for the 14-nm sample, an 8-nm thick layer of hBN was transferred onto it to study heat transport in the heterostructure, from which $G$ was extracted. All measurements were carried out in vacuum to prevent heat dissipation in air (or surrounding environment), which would be a source of error, leading to an overestimation of the heat conductivity of the suspended layers.

## Results

The 5-nm MoS$_2$ sample suspended over a 20-µm diameter hole and its corresponding Raman spectrum are shown in Fig. 2(a). For the 2LRT measurements, the heat was applied at the centre of the sample, while the Raman spectra were recorded at several points along the sample diameter. The change in the Raman A$_{1g}$ mode frequency was consequently converted to temperature, using χ as described previously; the resulting linescan, is shown in Fig. 2(b). The data provides information on the temperature rise at the centre of the sample (x= 0), the temperature decay along the sample and the temperature at the edges.

In order to extract the $k_{MoS_2}$ the spatial distribution of the temperature rise induced by the laser beam, the problem is reduced to a one-dimensional integral, which is evaluated numerically using the finite elements method (FEM) in COMSOL software. The solution for an arbitrary laser intensity distribution is specified for the case of a Gaussian beam (see Materials and Methods)[25]. The numerical solution is directly fitted to the experimental points shown in Fig. 2(b) for $P_{abs}$= 40 µW to obtain $k_{MoS_2}$.
Here $P_{abs}$ is determined experimentally by measuring the incident, reflected and transmitted power across the suspended layer. In this case, the best fit corresponds to $k_{MoS_2}$ = 12 ± 5 Wm$^{-1}$K$^{-1}$. It should also be noted that samples reach close to room temperature at the edges, confirming a good thermal contact and a high quality heat transfer process to the substrate.

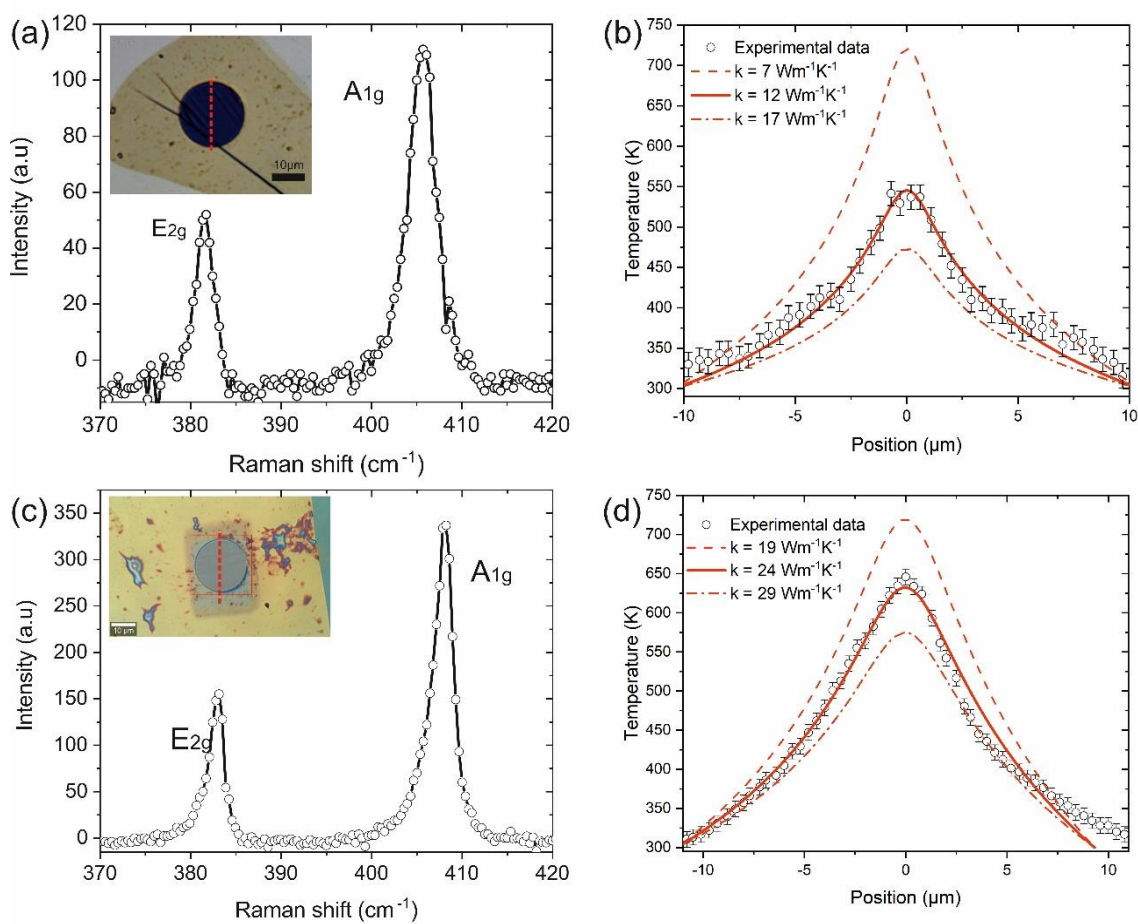

**Figure 2.** 2LRT of the single crystalline MoS$_2$ samples. Raman spectrum of the suspended (a) 5 nm- (c) 14 nm-thick MoS$_2$. Insets: optical images. Dashed lines indicate the scan direction of the laser probe. Temperature dependence as a function of the position on for (b) 5-nm thick sample with $P_{abs}$= 40 µW;

and (d) 14-nm thick sample with $P_{abs}$= 0.425 mW. The heating laser is focused at the center of the samples (equivalent to the Position equal 0).

The same experimental procedure was applied to the 14-nm thick MoS$_2$ sample suspended over 20-µm hole, as shown in Fig. 2(c). The temperature linescan in Fig. 2(d) demonstrates symmetric temperature distribution on the sample and a uniform thermal boundary where the sample was in contact with the substrate. The best fit to Fig. 2(d) for $P_{abs}$ = 0.425 mW yields a thermal conductivity $k_{MoS_2}$ = 24 ± 5 Wm$^{-1}$K$^{-1}$.

Finally, an 11-layer hBN flake was transferred onto the suspended MoS$_2$ using the dry transfer method [see inset in Fig. 3(a)]. The uniformity of both suspended layers is evidenced by Raman spectroscopy measurements, which highlight a uniform value of the frequency of the A$_{1g}$ mode and a constant difference between the frequencies of the A$_{1g}$ and E$_{2g}$ modes (Fig. S2). Because the intensity of the Raman peaks in the few-layer h-BN sample is weak compared to that in MoS$_2$, they are unsuitable for Raman thermometry techniques[19]. Indeed, it was not possible to obtain the temperature distribution in the hBN layer and the temperature could only be probed in the MoS$_2$, which we further used to calculate the effective $k$ of the heterostructure.

The temperature profile was obtained using the procedure implemented for the isolated MoS$_2$. Similar to the isolated MoS$_2$, the fit was obtained by introducing an effective $k$ for the MoS$_2$/hBN heterostructure. From the temperature profile in Fig 3(b), the $k$ was calculated to be 185 ± 20 Wm$^{-1}$K$^{-1}$. This value shows an almost eight-fold increase with respect to the isolated MoS$_2$ layer. Furthermore, up to 7.078 mW heating power was applied to the heterostructure, leading to a temperature increase of 550 K without any visible damage. In the case of isolated MoS$_2$, a laser power of just 0.425 mW leads to a temperature increase of 350 K. A temperature mapping for this sample was also performed showing a symmetric temperature distribution (Fig. S3b). These observations confirm that the hBN layer serves as an efficient in-plane heat spreader in the heterostructure.

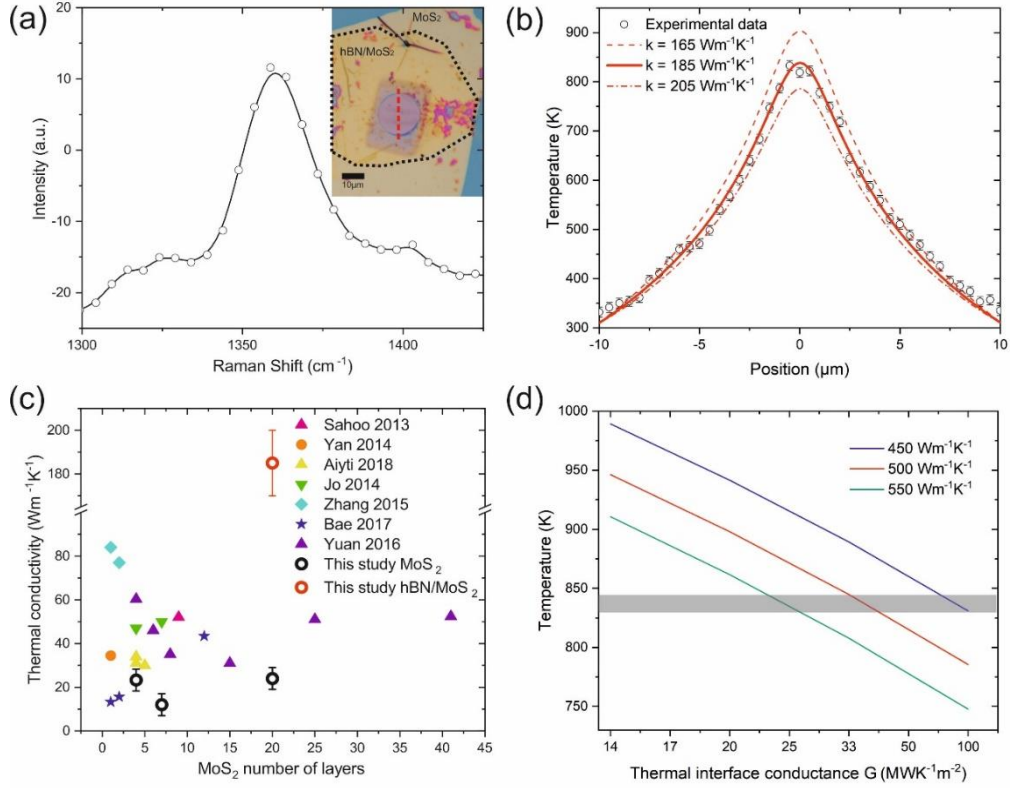

**Figure 3.** 2LRT of the MoS$_2$/hBN heterostructure. (a) Raman spectrum of the 8 nm thick hBN. The intensity is much smaller than that of MoS$_2$. Inset: optical image of the sample. (b) Temperature dependence as a function of the position at $P_{abs}$ = 7.078 mW. (c) Comparison of the measured values of thermal conductivity in MoS$_2$ in this study and the literature. (d) Temperature in the centre of the heterostructure as a function of MoS$_2$-hBN thermal interface conductance $G$, simulated for varying $k_{hBN}$.

**Discussion**

While the 1LRT is experimentally less challenging, it provides limited information on the thermal transport in the samples. In contrast, 2LRT provides a deeper picture of the different thermal transport processes in the system, in particular, it allows us to determine the spatial temperature distribution with high accuracy, which can provide information on the boundary thermal contact, the presence of hotspots and further resolve thermal anisotropy in the samples investigated. 1LRT is also not sensitive to sample quality, as Raman spectra are only taken at the center of the sample. The need of high-quality and clean samples becomes evident in 2LRT, as the polymer residues are visible in the linescans and maps. The temperature profile of an example MoS$_2$ layer transferred using PLLA polymer, and containing various residues, is shown in Fig S5. Such an inhomogeneity in less than ideal samples could easily be overlooked by 1LRT.

In Fig. 3(c) the thermal conductivity values obtained in this and previous works on crystalline MoS$_2$ are compared. Apart from the three samples described above a fourth one, a 3-nm thick MoS$_2$ layer suspended over the 10-µm hole was measured using 1LRT. Following the formalism developed previously [3,11] we obtain $k_{MoS_2}$ = 23 ± 5 Wm$^{-1}$K$^{-1}$ (see Supplementary Information). In our work, both 1LRT and 2LRT show consistent values of $k_{MoS_2}$ between 12 and 24 Wm$^{-1}$K$^{-1}$. However, these values appear to be slightly lower than many of the previously reported, which we attribute to differences in the experimental methods. For instance, the higher values reported in ref [4] are likely due to the fact that the measurements were performed in air. As explained in the experimental details section, all of the measurements in this work were performed in vacuum, which is crucial to extract the intrinsic $k$ of the 2D materials, as the heat conduction to air constitutes a significant dissipation channel and can be treated as a source of systematic error[15]. Even though we would expect some temperature dependence of $k_{MoS2}$, very good fits to the experimental data are obtained for constant $k_{MoS2}$. This assumption is shown to be valid within the quoted $k_{MoS2}$ uncertainty, as shown in Fig. 2 (b) and Fig. 2(d), where the expected temperature profiles are plotted for selected $k_{MoS2}$.

In order to extract $G$ between the MoS$_2$ and hBN we refer to the literature for the values of $k_{hBN}$. The bulk $k_{hBN}$ was previously measured to be of approximately 400 Wm$^{-1}$K$^{-1}$ and strongly dependent on the crystalline quality of the samples[26]. For single- and few-layer hBN, same as for the other 2D materials, the large disparity of $k$ values can be found in the literature. The thermal conductivity in single-crystalline hBN was reported as 751 ± 340, 646 ± 242, and 602 ±247 Wm$^{-1}$K$^{-1}$ at room temperature for one, two and three layers, respectively using 1LRT[21]. There, the absorption coefficient was directly measured using optical microscopy in transmission mode. The absorbance of the single hBN layer was ~0.5% for the 514 nm wavelength. Taking into account the large bandgap (~6 eV) of high-quality hBN very weak absorption is expected, however, defect states can significantly increase the absorption in the visible range. The $k_{hBN}$ of the 5- and 11-layer sample was determined to be 250 and 360 Wm$^{-1}$K$^{-1}$, respectively, using a microbridge device with built-in resistance thermometers[19]. However, in that work the presence of polymer residues on the sample surface was found to strongly affect the thermal conductivity, especially for the 5-layer sample.

Assuming an (infinitely) large $G$ between the MoS$_2$ and hBN, the temperature of MoS$_2$ and hBN would be approximately equal at the heatsink and the obtained $k$ from Fig. 3(b) would derive from the contributions of MoS$_2$ and hBN in a parallel configuration. Using the latter approximation, and knowing $k_{MoS2}$ = 24 Wm$^{-1}$K$^{-1}$, we can estimate $k_{hBN}$ ~ 460 Wm$^{-1}$K$^{-1}$, which is of the same order of magnitude as the bulk. However, once a finite $G$ is introduced, the temperature rise at the centre of the MoS$_2$ will be larger than expected by the simple parallel model and the temperature profile will be modified.

The value of $G$ for different $k_{hBN}$ can be more precisely estimated using COMSOL. For this purpose, we simulated the temperature rise at the hotspot with $P_{abs}$ = 7.078 mW, probed at the MoS$_2$ [Fig. 4 (c)]. The value of $k_{hBN}$ was varied between 450 and 550 Wm$^{-1}$K$^{-1}$, as previously reported, while $k_{MoS2}$ was again set to our experimental result, i.e. $k_{MoS2}$ = 24 Wm$^{-1}$K$^{-1}$. As observed in Fig. 3(d), $G$ ~ 75-100 MW·m$^{-2}$K$^{-1}$ is obtained with $k_{hBN}$ ~ 450 Wm$^{-1}$K$^{-1}$. For comparison, $G$ between graphene and hBN was reported to have a value of 52.2±2.1 MW·m$^{-2}$K$^{-1}$ [22]. If we assume higher values of $k_{hBN}$ between 500 and 550 Wm$^{-1}$K$^{-1}$ it will result in $G$ of 35 and 25 MW·m$^{-2}$K$^{-1}$, however such large $k_{hBN}$ are not expected and have been only reported in very thin layers of hBN.

This represents a dramatic increase in $G$ compared to previous reports with hBN and other interfaces. For MoS$_2$ layers on the typical CMOS substrates, such as SiO$_2$ or AlN, $G$ ~ 2 - 15 MW·m$^{-2}$K$^{-1}$ was measured [27,28]. These very low values mean that heat dissipation to the substrate might not be sufficient to assure correct device operation. Early measurements of thermal conductance of the MoS$_2$-hBN interface with 1LRT on a supported sample yielded the room-temperature value of 17.0±0.4 MW·m$^{-2}$K$^{-1}$. Also, a maximum theoretical limit of 26.4 MW·m$^{-2}$K$^{-1}$ was predicted using non-equilibrium Green's function, a value that is significantly lower than the one obtained in our work.

Finally, a recent work reported $G$ ~70 MW·m$^{-2}$K$^{-1}$ in MoS$_2$ fully-encapsulated with hBN (i.e. hBN/MoS$_2$/hBN) in both supported and suspended samples using 1LRT [29]. For the hBN/MoS$_2$ heterostructure, the estimated thermal conductance was only 23.8 MW·m$^{-2}$K$^{-1}$. Surprisingly, the absorbance 10.6 and 16.8 nm thick hBN was measured to be 16.7 and 25%, respectively, indicating the presence of either defects or contamination or both. This might represent a hint why the full encapsulation was needed in order to evacuate heat efficiently.

In conclusion, both of these prior works used Raman spectroscopy to study similar systems, but their contradictory results confirm the pitfalls of thermal measurements in 2D materials and how the quality of the samples and the experimental conditions may influence the estimated values of the thermal conductivity. Our results show significantly higher interfacial thermal conductance between the $MoS_2$ and hBN than any other substrate. The superior heat dissipation can be achieved with only one few-layer flake of hBN, which further confirms that a clean interface is essential for the high interface conductance which enables efficient heat transfer from the $MoS_2$ to the hBN layer.

**Conclusions**

We have studied the in-plane thermal conductivity of a few-layer $MoS_2$ and a $MoS_2$/hBN heterostructure using non-contact, Raman-based techniques. 2LRT has been successfully implemented for the first time to measure single-crystalline 2D materials and their heterostructures. We have measured $k_{MoS2}$, and showed that it varies between 12 and 24 $Wm^{-1}K^{-1}$ for the few-layer samples, in agreement with previous studies.

We showed that using mechanical exfoliation and dry transfer, is feasible to fabricate high-quality, suspended samples. The high quality of the suspended heterostructure is confirmed by the determination of its effective $k$. In particular, we show an eight-fold increase in $k$, and more importantly, that it is possible to dissipate in the heterostructure a laser power that is at least 10 times higher than in isolated $MoS_2$ with no damage to the material. The systematic measurements of the $MoS_2$/hBN heterostructure have demonstrated the concept of this material combination as an excellent interface to dissipate heat efficiently, opening the possibility to measure other types of 2D layered material interfaces and to determine with high accuracy their $k$ and $G$, eliminating the error arising from the presence of a bulk substrate. These results prove the potential of integrating few-layer hBN onto $MoS_2$-based devices for efficient in-plane heat spreading.

**Materials and methods**

*Raman characterisation*

Spectra shown in Figs. 3(d), S2 and S3 were obtained using a Witec Alpha300 R confocal Raman spectrometer and the 488 nm laser line.

All the other spectra were taken using a Horiba T64000 Raman spectrometer and the 532 nm laser line.

*Raman thermometry*

All the Raman thermometry measurements were performed in a Linkam temperature-controlled vacuum stage (THMS350V) under vacuum ($5 \times 10^{-3}$ mbar) at room temperature. Horiba T64000 Raman spectrometer and a 532 nm laser (Cobolt) were used to obtain the spectra.

In 1LRT the laser beam was focused on the sample with the microscope objective (50× and NA = 0.55) acting as a Gaussian heat source with a waist size of about 1 μm. The absorbed power, $P_{abs}$, was measured for each sample as the difference between the incident and the transmitted plus reflected laser power. The powers were measured with a calibrated system based on cube nonpolarizing beam splitters, i.e., no assumptions are made of the sample optical absorption. The error of absorbed power measurements was of $\Delta P_{abs} = 5\%$.

In the 2LRT experiments light emitted by the fibre-coupled continuous wave laser operating at 405 nm (Cobolt) was focused onto the sample from the bottom by a long working distance microscope objective (50x and NA = 0.55) acting as a Gaussian heat source with a waist size of about 1 µm. The absorbed power $P_{abs}$ is measured on site for each sample as the difference between incident and transmitted plus reflected light intensities probed by a calibrated system based on a cube non-polarising beam splitter with an error of $\Delta P_{abs} = 5\%$.

The probe laser beam with wavelength 532 nm is focused on the sample from the top, as explained in the previous section, and works as a temperature probe. To minimise the influence of the probe laser beam on the measurement, its power is set below the value (typically <10% of the heating laser power) which results in a measurable temperature rise.

*FEM simulations of 2LRT*

The spatial distribution of the temperature rise induced by a laser beam absorbed in a solid is reduced to a one-dimensional integral which is evaluated numerically. The solution for a general laser intensity distribution is specialized to the case of a Gaussian beam[25].
In order to extract the thermal conductivity *k* we need to solve Fourier's heat equation in the steady state:

$$\frac{P_{abs}}{A} = -k\nabla T$$

where *A* is the cross-sectional area of the heat flux, *k* is the thermal conductivity, and *T* is the temperature. Assuming Gaussian laser beam distribution:

$$P(r) = P_0 \exp(-\frac{2r^2}{w_0^2})$$

where *r* is the distance from the membrane center and $P_0$ and $w_0$ are the Gaussian beam amplitude and waist size, respectively. The relationship between the total absorbed power $P_{abs}$ and $P_0$ is:

$$P_0 = 2P_{abs}/(\pi w_0^2)$$

Because of the symmetry of the membrane and its isotropic in-plane thermal conductivity, the model was simplified to a 2D stationary heat flow study.

In case of one laser Raman thermometry temperature $T_p$ probed at the sample can be approximated by the formula:

$$T_p \approx \frac{\int_0^R T(r)P(r)r dr}{\int_0^R P(r)r dr}$$

where *T(r)* is the temperature distribution obtained from the FEM simulation and *R* is the membrane radius.

The simulations of 2LRT have been made using a FEM commercial software (COMSOL). We have assumed an effective medium model which simulates the temperature spatial profile over the circular membranes volume aiming to mimic the two-laser Raman thermometry experiment. As input parameters we use the absorbed power $P_{abs}$ (distributed in a Gaussian profile, which can be shifted from the centre of the membrane by an amount parametrized as $x_0$), the effective thickness and radius of the membrane (the temperature of the edges of the membrane, i.e., where the $MoS_2$ contacts the substrate, is kept to be 300K), a temperature-independent thermal conductivity ($k_0$) and the relative spatial shift between the centre of the heating laser and the probe laser ($y_0$). The latter means that, when doing the line profile, the probe laser does not exactly pass over the heating laser, the minimum distance between them being $y_0$.

We have simulated spatial temperature distributions $T(r)$ for a wide range of values for $x_0$, $k_0$ and $y_0$ (having the other parameters fixed to the values measured experimentally) with the objective of minimizing the difference between the experimental and simulated temperature curves. The figures reported afterwards reflect the results of this study, putting a special focus on the results obtained for the set of parameters that generate a simulated temperature profile that fits best to the experimental results.

**Acknowledgments**

This work was partially funded by the European Union under the H2020 FET-OPEN NANOPOLY (GA 289061) and Spanish Ministry of Science projects SIP (PGC2018-101743-B-I00), ADAGIO (PGC2018-094490-B-C22), 2DTecBio (FIS2017-85787-R) and 2DENGINE (PID2019-111773RB-I00/AEI/10.13039/501100011033). EDC acknowledges the Spanish Ministry of Science for the Juan de la Cierva Fellowship (JC-2015-25201) and the Ramon y Cajal fellowship (RYC2019-027879-I). DNU and JFS acknowledge the Ramón y Cajal fellowships RYC2014-15392 and RYC2019-028368-I/ AEI / 10.13039/501100011033. The Catalan Institute of Nanoscience and Nanotechnology (ICN2) is funded by the CERCA program/Generalitat de Catalunya, and is supported by the Severo Ochoa program from Spanish MINECO (Grant No. SEV-2017-0706).